\documentclass{elsart}
\usepackage{amsmath}
\usepackage{graphicx}
\usepackage{subfigure}
\usepackage{calc}
\usepackage{color}
\usepackage{amssymb}
\begin{document}
\begin{frontmatter}
\title{Robustness of a Cellular Automata Model for the HIV Infection\thanksref{agradecimentos}}
\author[ufrpe]{P. H. Figueir\^{e}do\corauthref{phf},}
\ead{phugof@df.ufrpe.br}
\corauth[phf]{phugo@df.ufrpe.br} 
\author[ufpe]{S.\ Coutinho,}
\ead{sergio@lftc.ufpe.br}
\author[ufpe]{R.\ M.\ Zorzenon dos Santos}
\ead{zorzenon@df.ufpe.br}
\address[ufrpe]{Departamento de F\'{\i}sica,
Universidade Federal Rural de Pernambuco, Dois Irm\~aos, CEP 52171-900, Recife,
Pernambuco, Brazil.}
\address[ufpe]{Laborat\'orio de F\'{\i}sica Te\'orica e Computacional, Departamento de F\'{\i}sica,
Universidade Federal de Pernambuco, Cidade Universit\'aria, CEP 50670-901, Recife,
Pernambuco, Brazil.}
\date{\today}
\thanks[agradecimentos]{This work was partially supported  by CNPq and CAPES (Brazilian federal
agencies) and by FACEPE (Pernambuco state agency) under the grant
PRONEX/FACEPE EDT 0012-05.03/04. PHF would like thank G.\ Camelo-Neto for some useful comments.}
\begin{abstract}

An investigation was conducted to study the robustness of the
results obtained from the cellular automata model which describes
the spread of the HIV infection within lymphoid tissues \cite
{PRL/RMZS-SC:2001}. The analysis focussed on the dynamic behavior of
the model when defined in lattices with different symmetries and
dimensionalities. The results illustrated that the three-phase
dynamics of the planar models suffered minor changes in relation to
lattice symmetry variations and, while differences were observed
regarding dimensionality changes, qualitative behavior was
preserved. A further investigation was conducted into primary
infection and sensitiveness of the latency period to variations of
the model's stochastic parameters over wide ranging values. The
variables characterizing primary infection and the latency period
exhibited power-law behavior when the stochastic parameters varied
over a few orders of magnitude. The power-law exponents were
approximately the same when lattice symmetry varied, but there was a
significant variation when dimensionality changed from two to three.
The dynamics of the three-dimensional model was also shown to be
insensitive to variations of the deterministic parameters related to
cell resistance to the infection, and the necessary time lag to
mount the specific immune response to HIV variants. The robustness
of the model demonstrated in this work reinforce that its basic
hypothesis are consistent with the three-stage dynamic of the HIV
infection observed in patients.
\end{abstract}

\maketitle

\end{frontmatter}

\section{Introduction}\label{section1}

Since the first cases of {\it Acquired Immunodeficiency Syndrome}
(AIDS) were notified during the early 1980s, considerable advances
have been made in understanding the dynamics of the infection
generated by the {\it Human Deficiency Virus} (HIV). What
differentiates HIV from many other viruses is its ability to remain
in the system after primary infection. Although the infectious
process varies from patient to patient, a typical infection pattern
has been observed amongst patients \cite{NEJM/PGF:1993,SC/JMC:1995}
and exhibits three phases: primary infection, the latency period and
the onset of AIDS. Primary infection is characterized by a broad
viral dissemination, which declines markedly (without eliminating
the virus)in weeks to months after the emergence of the first
HIV-specific immune response \cite{NEJM/DM_et.al:1991}. This rapid
decrease in the plasma viremia titer is followed by the
\textit{clinical latency} period, which may vary from 2 to 10 years
or more, with a very low viral burden. During the clinical latency
period, although many patients are usually asymptomatic, all suffer
a gradual performance deterioration of the immune systems,
manifested by the decrease of $CD4^+$ T cell counts
\cite{NEJM/PGF:1993} to very low concentrations. The onset of AIDS
is associated to achieving very low thresholds of $CD4^+$ T cell
counts corresponding to 20-35\%\ of the regular values of a normal
individual. In the absence of treatment, the outcome of this stage
is death by opportunistic diseases.

The main target of the HIV is the $CD4^+$ T cell. The viral envelop
protein binds to some specific cell receptor, thus enabling the
virus to deposit its genetic material into the cell. Once inside, it
uses the host cell machinery to make copies of its viral DNA in the
same manner as other retroviruses \cite{nowak00}. When the host
cells are able to produce both the DNA and the viral protein
envelop, they are able to release new viral particles into the
extra-cellular environment. Replication of HIV is extremely fast and
mutation rates are very high. It is estimated that during the HIV
reproduction process, two generations are produced per week with one
mutation per generation \cite{nowak00}. Part of the new strains
would be recognized by previously developed immune responses, but
part of the variants would not be recognized and would activate new
specific immune responses to them. Since the virus is not
eliminated, the process of continuously activating a new specific
immune response is maintained indefinitely, thus placing stress on
the immune system. The ultimate manifestation of the effects of this
stress is a decline in T cell counts. Several mechanisms have been
proposed to explain the decline of T-cell counts
\cite{NEJM/PGF:1993,SC/W:1993} the aggregation of infected and
non-infected cells in the lymphoid tissues (syncytia formation), the
destruction of memory-T cells, programmed cell death associated to
lymphocyte activation instead of cell proliferation, and high viral
replication and mutation rates. However, despite all efforts to
determine the cause of this decline during recent decades, the
question still remains open-ended.

Over the past two decades many mathematical models based on
differential equation approaches have been proposed in order to
study the different aspects involved in the development of the HIV
infection \cite{MB/PN_etal:2000}. Most of these models however,
describe the evolution of the virus and cell populations in a
compartmentalized manner, and do not take into account the spatial
localization necessary to mount the specific immune responses
~\cite{nowak00}. Although this kind of approach has contributed to
the understanding of different aspects of HIV infection dynamics, it
has not been able to reproduce the entire three-stage dynamics
observed in infected patients. Depending on the adopted approach,
the model only adequately describes either the early (stochastic
models) or later stages (deterministic models) of the
dynamics~\cite{MB/PN_etal:2000}. There is also a chance it may
describe the two time scales using different sets of parameters for
different stages~\cite{AMS/DEK:1996}. Discrete approaches based on
binary cellular automata have also been
proposed~\cite{TB/P&M&R:2000} to describe the interactions amongst
$CD4^+$ and $CD8^+$ T cells, macrophages and viral particles. These
models reproduce the different attractors of the
dynamics~\cite{JSP/P&S:1990} or the early stages of the infection
~\cite{TB/P&M&R:2000} but not its entire course. Nowadays, spatially
structured models are being recognized as a step forward in
understanding the dynamics of virus infection processes, in which
the influence of local dispersal of the virus and virus-target
 cells is relevant for disease persistence in vivo. This is the case of
 the HIV infection~\cite{funk2005,strain02a}.

Recently, two of the authors have proposed a two-dimensional
stochastic cellular automata (CA) model in order to describe the
spread of the HIV infection \cite{PRL/RMZS-SC:2001}, and which takes
into account the spatial localization of target cells (T cells) that
occurs in the lymphoid tissues to mount the specific immune
response. In the case of the HIV infection it is this localization
that contributes to the spread of infection throughout the cells.
The model illustrates that the combination of the timescale involved
in the immune response of any (healthy) individual with fast
replication and high mutation rates of the HIV, together with the
spatial localization generated in the lymphoid tissue, may go some
way to explaining the three-stage dynamics observed in experimental
findings \cite{NEJM/PGF:1993,SC/JMC:1995}. By reproducing the two
timescales observed in the experimental data (concerning T cell
counts and viremia titer) the model permits the possible mechanisms
underlying the dynamics of this infectious process to be
investigated. The results suggest that the slow timescale is
associated to the agglutination of infected cells in structures,
which may compromise the entire tissue and trap healthy cells,
leading to the onset of AIDS. All these structures may be associated
to syncytia formation. Syncytia are structures, which are formed
when infected blood is mixed with healthy blood, in \textit{in
vitro} experiments. Therefore, results suggest that what is observed
\textit{in vitro} may also occur \textit{in vivo}.

The cellular automata model proposed in reference
\cite{PRL/RMZS-SC:2001} is defined on a two-dimensional square
lattice so as to mimic the structure of  lymphoid tissues where cell
interactions take place. In the case of HIV infection, lymph nodes
play a major role within lymphoid tissues. The lymph nodes are small
organs exhibiting a bean shape. From electromicrography images
\cite{hood} the region where the interactions between the different
cells and the viruses take place, has a fractal structure similar to
that of a sponge. The cells take in the order of hours to cross the
lymphoid tissue and find the right place to interact within the
porous structure. In the model, the range of interaction between the
cells is approximated by a surface using a square lattice. Despite
the approximation and remarkable accuracy of the model in describing
the two timescales and the three-stage dynamics, the question
concerning the relevance of the lattice geometry as well as its
dimensionality (2-D, 3-D or fractal) remains an important issue to
be investigated. This issue was taken up and addressed by
Figueir\^edo \cite{Msc.Thesis:2002} whose results are for the first
time reported in this paper. Recently, Omerond \cite{omerod04}
conducted a similar investigation into how different tilings of
square and cubic lattices, and definitions of the nearest neighbors
qualitatively affect the dynamics of the system, without exploiting
the robustness. However, while being a more qualitative
investigation, it was less detailed than the current study.

Below, further results will also be presented concerning the
robustness  of the dynamics of the model when varying the stochastic
parameters of the model. To be more precise, studies were undertaken
into the characteristic changes of the primary infection, due to
variations of the initial HIV concentration for both the two and
three-dimensional models. Moreover, an analysis was carried out of
the time-scaling behavior of the clinical latency period based on
the probability that newly infected cells enter the system
throughout time, by different lattices. In order to complete the
analysis of the cubic lattice, an investigation was also conducted
into the behavior of the average latency period as a function of the
parameters that governs the intensity and the time delay to mount a
specific immune response.

This paper is organized as follows: In section \ref{section2}, the
model  is presented together with a discussion of the modifications
introduced in order to deal with the triangular and cubic lattices.
In section \ref{section3}, the obtained results are presented and
discussed, and the conclusions are presented in section
\ref{section4}.

\section{The Cellular Automata Model}\label{section2}

In the CA model~\cite{PRL/RMZS-SC:2001} a target cell (T cells or
monocytes) is associated to each site of a square lattice. Each cell
is represented by a four-state automaton corresponding to different
states of the cell: \emph{healthy}, \emph{infected-A},
\emph{infected-B} and \emph{dead}.  The infected-A state corresponds
to a virus-producing cell, which promotes the spread of the
infection during a time interval $\tau$ without any suppression.
$\tau$ corresponds to the period of time necessary for the immune
system to mount a specific immune response to a given virus strain.
A virus-producing cell that was already recognized by the immune
system is then represented by the infected-B state. Although
infected-B cells are less effective than infected-A in contaminating
healthy cells, a high concentration in a given region may
disseminate the infection throughout the nearby uninfected cells.
The infected-B cells eliminated by the cytotoxic mechanisms of the
specific immune response or the cytophatic effects of virus
replication are represented by the dead state. The high ability of
the immune system to replenish depleted cells is taken into account
by converting a fraction $p_{\textsf{repl}}$ of the dead cells into
new cells. Such new cells mimic the flux of incoming cells from
other compartments. Most of the new incoming cells would be healthy,
but a (small) fraction of them ($p_{\textsf{infec}}$) would be
infected-A, corresponding to virus-producing cells coming from other
compartments.

Simulations were carried on a $L \times L$ square lattice starting
from an initial configuration composed of healthy cells, with small fraction,
$p_{\textsf{HIV}}$, of infected-A cells randomly distributed among the healthy cells,
representing the initial contamination by HIV. Periodic boundary
conditions were considered and each time step corresponded to the parallel
updating of the entire lattice according to the following rules:
\begin{description}
\item [1)] A healthy cell becomes an infected-A if it has at least one
infected-A cell amongst its nearest neighbors, or at least $R=4$ neighbors
in infected-B state. Otherwise it remains healthy.
\item [2)] An infected-A cell remains in this state during $\tau$ time
steps, after which it becomes infected-B.
\item [3)] An infected-B cell becomes a dead cell in the next time step.
\item [4)] A dead cell is replaced by a healthy cell with probability
$(1-p_{\textsf{infec}})\,p_{\textsf{repl}}$, or by an infected-A
cell with probability $p_{\textsf{repl}}\,p_{\textsf{infec}}$,
otherwise it continues in the dead state with probability
$(1-p_{\textsf{repl}})$.
\end{description}

Some of the parameter values chosen are based on experimental
findings.  For example, one in $10^4$ to $10^5$ cells in the
peripheral blood of infected patients expresses viral proteins
\cite{SC/AS.F:1996} during the latency period, and
$p_{\textsf{infec}}$ values are brought about by this finding. The
probability of replacing dead cells by new incoming cells
($p_{\textsf{repl}}$) could vary between zero and one and may also
vary from patient to patient. In the original model it was assumed
that $p_{\textsf{repl}}=0.99$ corresponds to a high replenishment
capacity of the immune system, which in this case is not affected by
the HIV infection, and is contrary to certain assumptions cited in
the literature. Such a replenishment concerns to the population of
cells that was recruited or committed do participate in the immune
response. Finally, the small fraction $p_{\textsf{HIV}}$ of infected cells randomly
distributed in the initial configuration was chosen according to
experimental findings, which indicates that one in $10^2$ to $10^3$
cells harbor viral DNA during primary infection
\cite{SC/S_et.al:1989,AIM/S_et.al:1990}.

According to the updated rules adopted since the beginning,
infection  is disseminated in a deterministic manner, driven by
cell-cell contact, but eventually newly infected cells are
introduced to the system due to stochastic rule 4. Rule 4 accounts
for all mechanisms governing the (re)infection process, such as the
presence of quiescent infected cells or infected cells coming to
other compartments.

In order to investigate the role of lattice symmetry and its
dimensionality on CA model dynamics, studies were conducted for the
models on the triangular (2D) and cubic (3D) lattices, to compare
results with those previously obtained using the square lattice. For
such cases, few changes should be made to the rules regarding the
neighborhood, as described above:
\begin{itemize}
\item Triangular lattice: \\
Each site possessed six neighbors, and $R=3$ was assumed as the
minimum number of infected-B neighboring cells required to
disseminate the infection, i.e., following the same criteria adopted
for the square lattice, this number would correspond to half the
number of neighbor sites.
\item Cubic lattice: \\
Each site possessed twenty six neighbors, hence $R=13$ was assumed
as the minimum number of infected-B neighboring cells required to to
infect a healthy cell.
\end{itemize}

To test the robustness of the model's dynamic behavior regarding the
variation of stochastic parameters, the position and fraction of the
infected cells at primary infection peak were measured, together
with the duration of the latency period when varying respectively,
$p_{\textsf{HIV}}$ and $p_{\textsf{infec}}$, over wide intervals.
The sensitivity of the latency period was also measured in the
three-dimensional model, when changing the deterministic parameters
$R$ and $\tau$. These results are presented and discussed throughout
the following section.

\section{Results and discussion} \label{section3}

\begin{figure}[h]
\begin{center}
{\includegraphics[width=16cm]{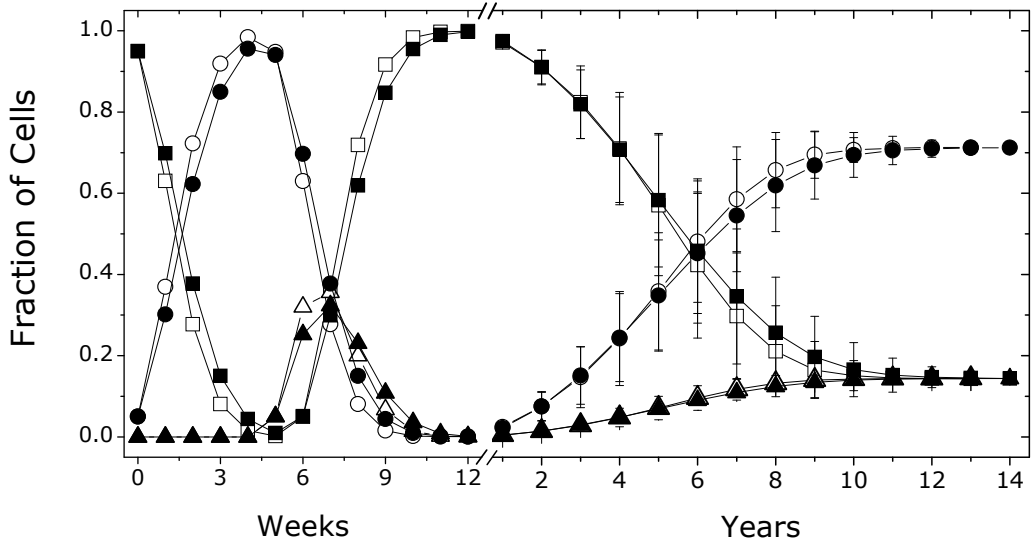}}
 \caption{Time evolution of the fractions of healthy (squares), infected-$A+B$
(circle) and dead (triangles) cells. Solid symbols correspond to the
results of the triangular lattice model ($R=3$), while open symbols
refer to the square lattice model ($R=4$).} \label{figure1}
\end{center}
\end{figure}

In order to analyze the effects of the changes in the symmetry of
the  lattice, an average three-stage pattern was obtained that
emerged from the CA model defined on the triangular and square
lattices for $L=900$ and the same set of parameters used in the
original model~\cite{PRL/RMZS-SC:2001}: $p_{\textsf{HIV}} = 0.05$,
$\tau = 4$, $p_{\textsf{repl}} =0.99$, $p_{\textsf{infec}} =
10^{-5}$. The results shown in figure~\ref{figure1} were an average
of 1000 samples (initial configurations) corresponding to different
individuals. Hereafter the error bars shown in the plots of all
figures indicate one-standard deviation.

As can be observed, there are no major quantitative changes in the
patterns obtained for the two lattices (square and
triangular), indicating that the three-stage dynamics is quite
independent of the lattice symmetry.
\begin{figure}[h]
\begin{center}
{\includegraphics[width=16cm]{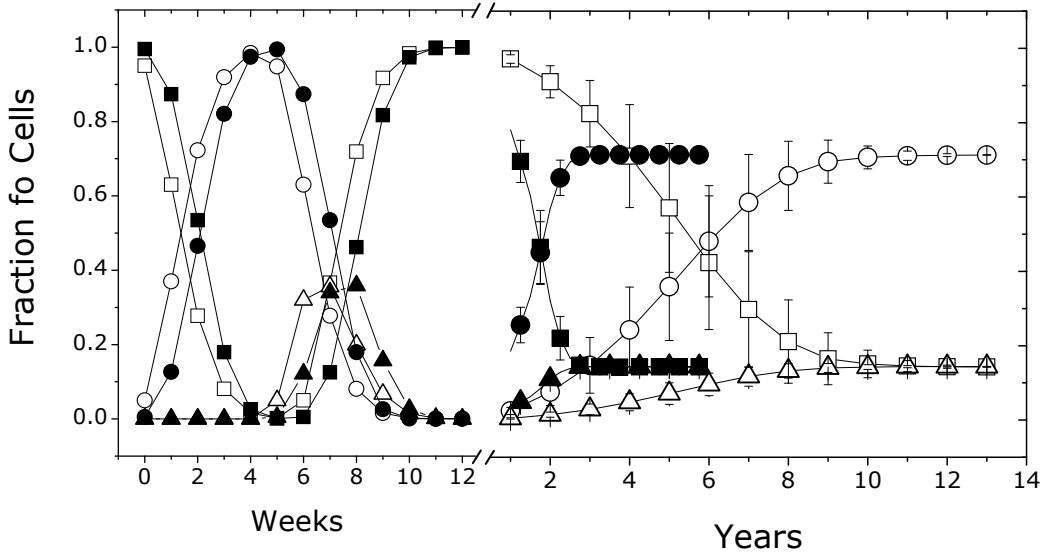}} \caption{Time evolution
of the fractions of healthy (squares), infected $A+B$ (circle) and
dead (triangles) cells. Solid symbols correspond to the results
obtained for the cubic lattice with $L=300$ and $R=13$ while open
symbols correspond to the square lattice using $L=900$ and $R=4$.
The parameters are those adopted to obtain Figure 1.}
\label{figure2}
\end{center}
\end{figure}
Figure~\ref{figure2} compares the results obtained from the model
defined on the cubic and square lattices. Except for the changes in
$L$ and $R$, all parameters were those adopted in ref~
\cite{PRL/RMZS-SC:2001}. It is worth observing that when the results
for different dimensions were compared, significant timescale
changes were observed. However, qualitative behavior remained the
same in both cases. The differences of the three-dimensional results
in relation to those of the two-dimensional were: a one week shift
in the peak of the primary infection phase, and a reduction in the
duration of the latency period. The duration of the latency period
is estimated for each patient by calculating the time necessary for
the fraction of healthy cells to achieve a threshold of $\sim 30\%$\, after
the primary infection. In this manner it was possible to estimate
the point that corresponds to the onset of AIDS
\cite{NEJM/PGF:1993}. The average length of the latency period (in
over $1000$ samples) was reduced from 7 (2-D) to 2 years(3-D). A
decrease in the error bars magnitude was also observed. This result
is easily understood when considering the number of neighbors in the
different lattices. While the cubic lattice possessed 26 nearest
neighbors, the square and triangular lattices possessed 8 and 6
neighbors, respectively. Hence, the large number of neighbors in the
3D case will favor a much faster dissemination of the infection,
when compared to the 2D lattices, thus reducing the latency period.
When considering the lymphoid tissue as a fractal object, according
to the results of this study, the fractal dimension should be closer
to two, thus validating the approximation of a two dimensional
lattice adopted in the original model~\cite{PRL/RMZS-SC:2001}.
\begin{figure}[h]
\begin{center}
{\includegraphics[width=12cm]{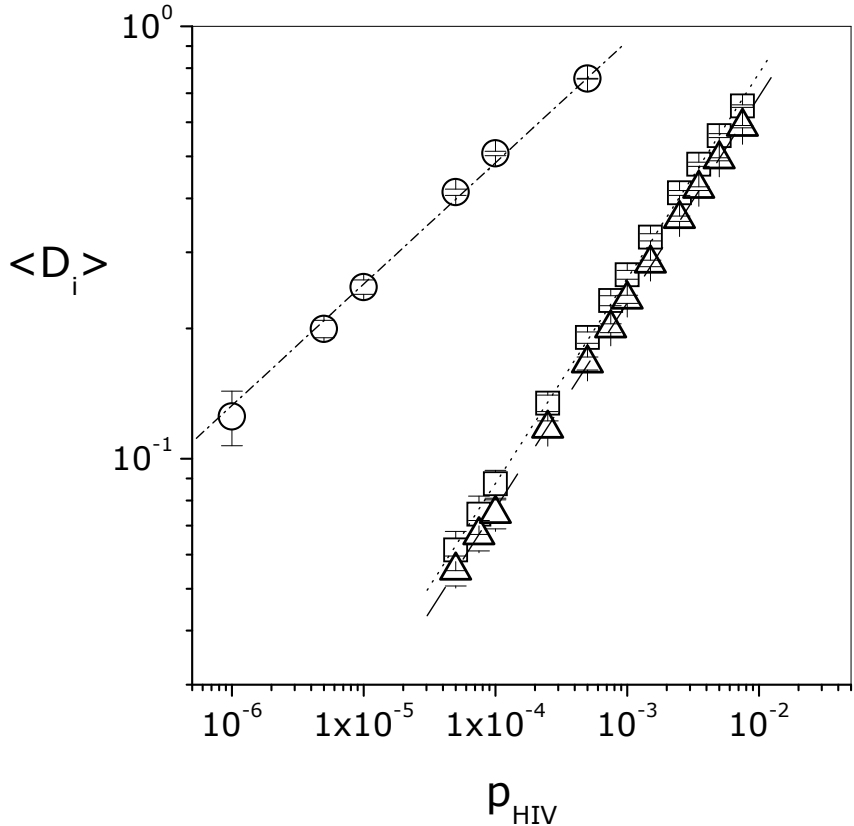}} \caption{Average maximum
fraction of infected cells during primary infection as a function of
$p_{\textsf{HIV}}$. The different plots correspond to: cubic
($\bigcirc$), square ($\square$) and triangular ($\bigtriangleup$)
lattices. The results were an average of over $1000$ samples the
error bars being of the order of the symbol size. Dashed lines
indicate the linear fitting.} \label{figure3}
\end{center}
\end{figure}
\begin{figure}[!h]
\begin{center}
{\includegraphics[width=12cm]{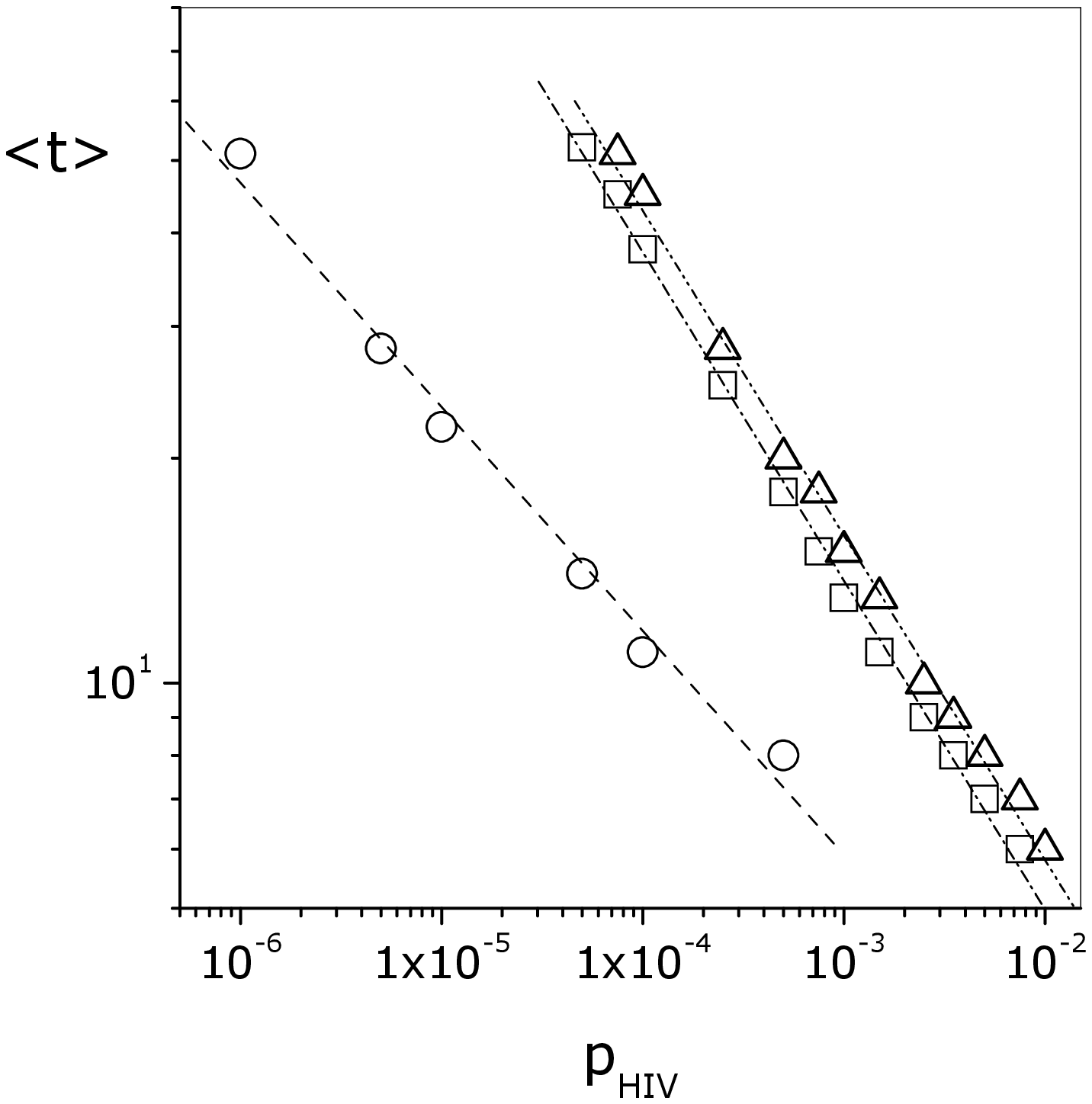}} \caption{Average
position of the maximum fraction of infected cells achieved during
primary infection ($<t>$) as a function of $p_{\textsf{HIV}}$.
Symbols represent the same lattices as described in
Figure~\ref{figure3} and the calculation of averages and error bars
also follows the same procedures as in the previous figure. Dashed
lines indicate the linear fitting.} \label{figure4}
\end{center}
\end{figure}
The robustness of the dynamics of the CA model during primary
infection  was also investigated for variations of the initial
concentration of infected cell ($p_{\textsf{HIV}}$) for 2D and 3D
versions of the model. Figure ~\ref{figure3} and ~\ref{figure4}
respectively, on a logarithmic scale, reveal the behavior of two
quantities that characterize the primary infection phase: the
average maximum concentration of infected cells $<D_{i}>$ and its
position $<t>$ (in time steps) as a function of $p_{\textsf{HIV}}$.
The small error bars indicate the robustness of the results
irrespective of the lattice dimensionality. Power-law behavior was
observed, indicating $<D_{i}> \sim (p_{\textsf{HIV}})^{\alpha}$ and
$<t> \sim (p_{\textsf{HIV}})^{\beta}$ with exponents $\alpha = 0.48
\pm 0.01$ for two-dimensional square and triangular lattices and
$\alpha = 0.29 \pm 0.01$ for the cubic. These values of $\alpha$ are
related to the average distance between first-nearest-neighbors
infected-A cells in the initial configuration, which by it turn scales with $1/D$.
Each infected-A cell of the initial configuration produces a pulse of infected cells, of
width ($\tau+1$) propagating in all directions. Whenever
such average distance is less than or equal to ($2\tau+1$),
the independent pulses achieve a maximum coverage of the
lattice, which corresponds to a maximum of $<D_{i}>$. The values obtained for
$\beta$ are similar for the two-dimensional lattices: $\beta = 0.44
\pm 0.01$, but a decrease was recorded of $0.30 \pm 0.02$ for the
cubic. In other words, the primary infection peak becomes broader
with the decrease of height by one order of magnitude while
$p_{\textsf{HIV}}$ varies over two decades for two-dimensional
lattices and three decades for three-dimensional. Other authors have
claimed that there exists a lack of robustness of the CA model when
describing primary infection ~\cite{strain02b}. However, the results
of this study assert the robustness of the results obtained with
respect to variations of the minimum concentration of infected cells
necessary to launch the infection process. The claim was based on an
unproven assumption of exponential growth of the viremia titer in
the very beginning of primary infection, and on speculative values
for the minimum amount of virus necessary to launch the
contamination of an individual, inferred from a few clinical cases.
This study illustrates that the same dynamics are launched for a
wide range of the initial concentration of infected cells.

\begin{figure}[h]
\begin{center}
{\includegraphics[width=12cm]{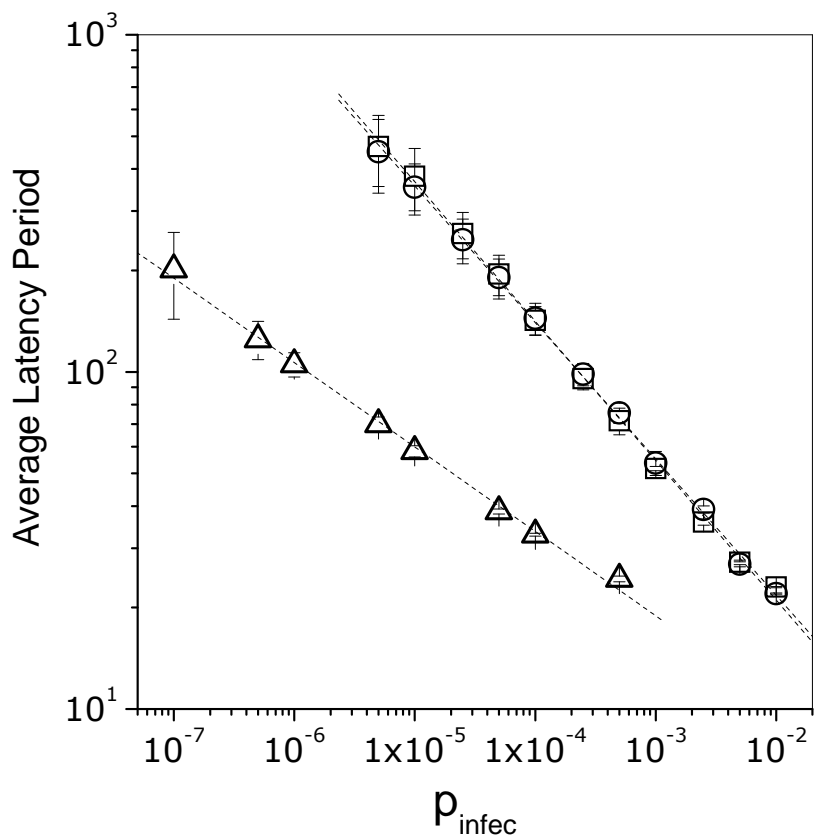}} \caption{Average
clinical latency period as a function of $p_{\textsf{infec}}$. The
different plots correspond to: cubic ($\bigcirc$), square
($\square$) and triangular ($\bigtriangleup$) lattices. Results were
an average of over 1000 samples. Dashed lines indicate the linear
fitting.} \label{figure5}
\end{center}
\end{figure}

The average latency period was also investigated as a function of
the probability of newly infected cells entering the system
($p_{\textsf{repl}} \times p_{\textsf{infec}}$). Since the
hypothesis was adopted that the function of the immune system is not
affected by the infection, and the replenishment of healthy cells by
the bone marrow is maintained constant  ($p_{\textsf{reg}}=0.99$),
we have plotted the average latency period solely as a function of
$p_{\textsf{infec}}$, as shown in figure ~\ref{figure5}. The
behavior of the average latency period is similar for both the
triangular and square lattices, but is different in the case of the
cubic lattice. The power law exponent $\gamma$ associated to the
average latency period is $0.40\pm 0.01$ for square and triangular
lattices, and $0.25\pm 0.01$ for the cubic.

\begin{figure}[h]
\begin{center}
\includegraphics[width=14cm]{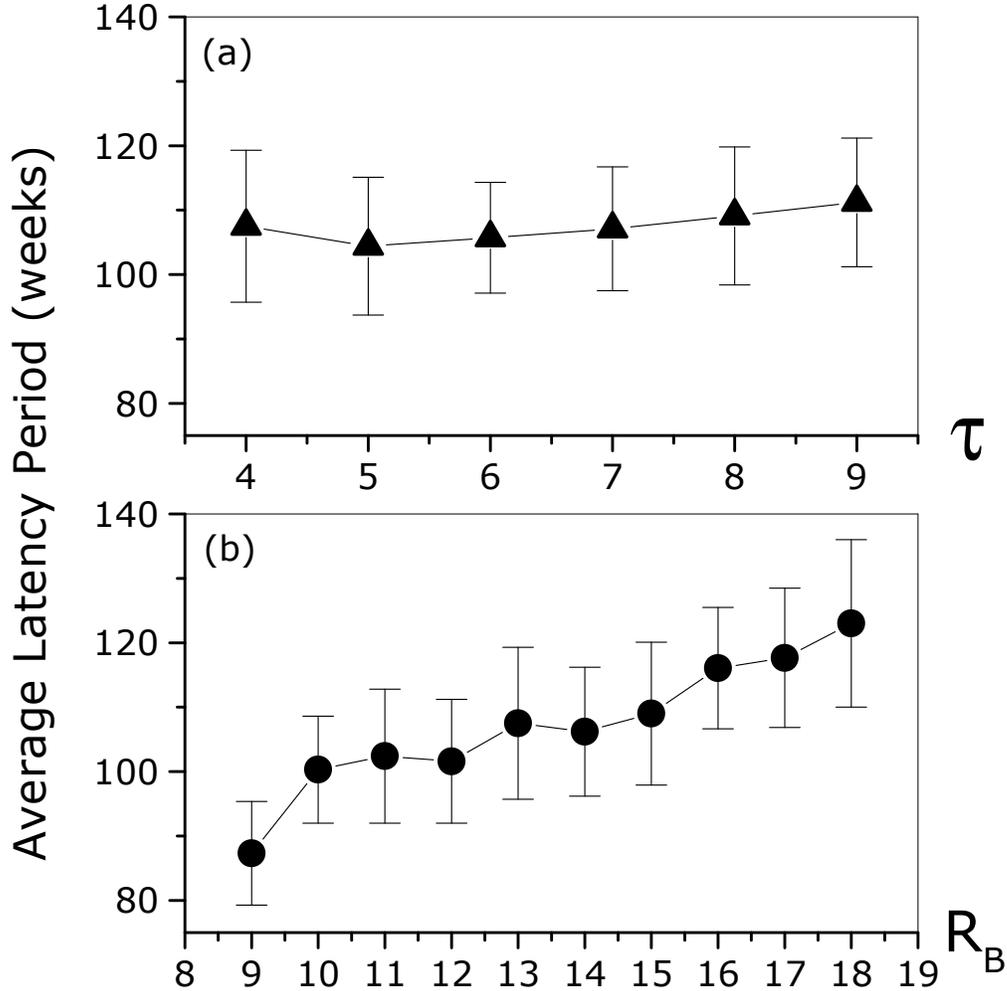}
\caption{Average clinical latency period for the cubic lattice
model: (a) as function of $\tau$; (b) as a function of $R$. Results
were averaged over 50 samples, using $L=300$.} \label{figure6}
\end{center}
\end{figure}

As previously mentioned, when the same set of parameters (except of
$L$ and $R$) are taken and the dynamical behavior exhibited by the
two and three-dimensional models is compared, a significant
reduction of the average latency period can be observed. Recently,
Solovey et al.\cite{solovey04} investigated how the variations on
the deterministic parameters change the dynamics of the model in the
square lattice. To complement the analysis concerning the
differences between the two and three dimensional models, this study
investigated the behavior of the average latency period in the
three-dimensional model when varying $R$ and $\tau$. The results in
figure~\ref{figure6}(a) indicate that, as in the case of the square
lattice~\cite{solovey04}, the average latency period is not
significantly altered due to variations in the time-lag parameter
$\tau$. However, when parameter $R$ varies from $9$ to $18$ an
increase in the latency period is observed to the order of 40 weeks,
as displayed in figure~\ref{figure6}(b). Such an increase
corresponds to a decrease in the probability of the occurrence of
infected cells and concomitant compact structures leading to long
latency periods, as would be expected. It can be noticed that both
plots of figure~\ref{figure6} exhibit large fluctuations. Such
fluctuations are related to the intrinsic width of the latency
period distribution and not with the number of simulations. In fact,
fluctuations are governed by the spatial- and \emph{time-dependent}
probability of the appearance of compact structures, and the latency
period is related to the average distance of such structures, which
becomes proportional to the inverse of the cubic ratio of the
concentration $\rho$ ($\sim \rho^{-1/3}$) for the three-dimensional
lattice.

\section{Conclusions}
\label{section4}

This paper has analyzed the robustness of the cellular automata
model  proposed in reference \cite{PRL/RMZS-SC:2001} to describe the
HIV infection with respect to changes in lattice symmetry and
dimensionality and in variations of the stochastic parameters of the
model. It has been shown that during the entire course of the
infection, the time evolution of the fractions of healthy, infected and
dead cells, are quantitatively reproduced irrespective of the
lattice symmetry in two dimensions. However, the average latency
period was found to be greatly reduced when the lattice dimension
changes from two to three, while the primary infection scale and the
overall qualitative behavior remains almost unchanged. These results
indicate that interactions within the lymph nodes occur on an
effective ``surface" with a fractal dimension close to two. This
therefore explains the success obtained with the two-dimensional
approximation used in the original model.

The robustness of the three-stage pattern was also observed when
varying the stochastic parameters associated to the minimum amount
of infected cells in the initial configuration necessary to launch
the infectious process ($p_{\textsf{HIV}}$) and to the mechanisms
responsible for either incoming newly infected cells or activating
quiescent infected cells ($p_{\textsf{infec}}$). The results
indicate quite a remarkable robustness of the three-scale dynamic
pattern following power-law behaviors over three orders of magnitude
of $p_{\textsf{HIV}}$. The observed width increase together with the
decrease of the primary infection peak, indicates that the lower the
initial concentration of infected cells, so the spread of the
infection within the tissue is slower and weaker, as would be
expected.

The behavior of the average latency period as a function of
$p_{\textsf{infec}}$ was also investigated for the three different
lattices (square, triangular and cubic). The results also exhibit
power-law behaviors over four (2D) to five (3D) decades: the lower
the $p_{\textsf{infec}}$ the greater the latency period, since the
probability of forming structures associated to the rapid evolution
of the onset of AIDS is reduced. In all cases studied, it was
observed that the lasting latency period depends on the formation of
spatial structures. Once these structures are formed they spreads
all over the lattice, thus compromising tissue and trapping healthy
cells, as observed in the square lattice
model~\cite{PRL/RMZS-SC:2001}. Such structures could be associated
to aggregates of infected cells observed {\it in vitro} experiments,
namely syncytia, and in the lymph nodes of infected patients.
Therefore, by means of the robustness analysis, the hypothesis can
be validated that the formation of such structures may be
responsible for the persistence of the virus in the system after
primary infection \cite{SC/JMC:1995,CTMI/Coffin:1992}. The
appearance of such aggregates, which naturally emerge from the
dynamics of the system in the automata cellular model, suggest that
a deeper biological investigation should be performed in order to
confirm the existence of these structures {\it in vivo} and their
role in the HIV infection.

\end{document}